\documentclass[twocolumn]{aastex701}

\begin{document}

\title{X-ray variability of SDSS J000532.84+200717.4: from a normal state to an X-ray weak state}

\author[orcid=0009-0007-7426-2758]{XiaoHui Yang}
\affiliation{Shenzhen Key Laboratory of Ultraintense Laser and Advanced Material Technology, Center for Intense Laser Application Technology, and College of Engineering Physics, Shenzhen Technology University, Shenzhen 518118, People's Republic of China}
\email{xiaohuiyang.astro@gmail.com}

\correspondingauthor{YanLi Ai}
\email{aiyanli@sztu.edu.cn}

\author[orcid=0000-0003-4897-4106]{YanLi Ai}
\affiliation{Shenzhen Key Laboratory of Ultraintense Laser and Advanced Material Technology, Center for Intense Laser Application Technology, and College of Engineering Physics, Shenzhen Technology University, Shenzhen 518118, People's Republic of China}
\email{aiyanli@sztu.edu.cn}

\author[orcid=0000-0002-4757-8622]{Liming Dou}
\affiliation{Department of Astronomy, School of Physics and Material Sciences, Guangzhou University, Guangzhou 510006, People’s Republic of China}
\email{doulm@gzhu.edu.cn}

\author[orcid=0000-0002-1517-6792]{Tinggui Wang}
\affiliation{Department of Astronomy, University of Science and Technology of China, Hefei, 230026, People’s Republic of China}
\email{tWang2014@ustc.edu.cn}

\author[orcid=0000-0002-2006-1615]{Chichuan Jin}
\affiliation{National Astronomical Observatories, Chinese Academy of Sciences, 20A Datun Road, Beijing 100101, People’s Republic of China}
\affiliation{School of Astronomy and Space Sciences, University of Chinese Academy of Sciences, 19A Yuquan Road, Beijing 100049, People’s Republic of China}
\email{ccjin@bao.ac.cn}

\author{Wenfeng Wen}
\affiliation{Shenzhen Key Laboratory of Ultraintense Laser and Advanced Material Technology, Center for Intense Laser Application Technology, and College of Engineering Physics, Shenzhen Technology University, Shenzhen 518118, People's Republic of China}
\email{2300411002@email.szu.edu.cn}

\author[orcid=0000-0003-4897-4106]{Xu Zhang}
\affiliation{Shenzhen Key Laboratory of Ultraintense Laser and Advanced Material Technology, Center for Intense Laser Application Technology, and College of Engineering Physics, Shenzhen Technology University, Shenzhen 518118, People's Republic of China}
\email{2410262039@stumail.sztu.edu.cn}

\author[orcid=0000-0002-0759-0504]{Yuming Fu}
\affiliation{Leiden Observatory, Leiden University, Einsteinweg 55, 2333 CC Leiden, The Netherlands}
\affiliation{Kapteyn Astronomical Institute, University of Groningen, P.O. Box 800, 9700 AV Groningen, The Netherlands}
\email{yfu@strw.leidenuniv.nl}

\author[orcid=0009-0007-5679-7724]{Jinhong Chen}
\affiliation{Department of Physics, University of Hong Kong, Pokfulam Road, Hong Kong, People's Republic of China}
\email{chenjh258@mail2.sysu.edu.cn}

\author[orcid=0000-0002-1517-6792]{Ning Jiang}
\affiliation{Department of Astronomy, University of Science and Technology of China, Hefei, 230026, People’s Republic of China}
\email{jnac@ustc.edu.cn}

\author[orcid=0000-0002-5310-3084]{Fukun Liu}
\affiliation{Department of Astronomy, School of Physics, Peking University, Beijing 100871, People's Republic of China}
\affiliation{Kavli Institute for Astronomy and Astrophysics, Peking University, Beijing 100871, People’s Republic of China}
\email{fkLiu@pku.edu.cn}

\begin{abstract}
We present a multi-epoch study of the extreme X-ray variability of the
type~1 quasar SDSS~J000532.84+200717.4 using archival observations from
\textit{XMM-Newton}, \textit{Swift}/XRT, \textit{EP-FXT}, and
\textit{ROSAT}, together with new optical spectroscopy and
multi-wavelength photometry. The 0.2--10~keV X-ray flux exhibits a
transition from a high state to a subsequent low state, declining by
more than an order of magnitude and placing the source in the X-ray--weak
regime ($\Delta\alpha_{\rm ox} \lesssim -0.3$). Significant variability
on timescales of days to weeks persists within the low state. In
contrast, the optical and mid-infrared emission remain stable over
decade-long timescales, while the UV continuum varies only mildly and
broadly tracks the X-ray evolution. Multi-epoch optical spectroscopy shows no significant long-term changes
in either the continuum shape or the broad emission-line profiles. The
\ion{Mg}{2} emission is relatively weak compared with typical quasars,
suggesting similarities to weak-line quasars. The pronounced wavelength-dependent
variability indicates that the accretion disk remains largely intact
while the X-ray emission undergoes dramatic changes. The spectral
hardening in the low state and the viability of ionized partial-covering
models are consistent with variable, largely dust-free absorbing gas,
possibly associated with clumpy inner disk winds, although intrinsic
coronal variations cannot be excluded. SDSS~J0005+200717.4 therefore
provides evidence that extreme X-ray weakness can arise as a transient
phase in otherwise normal quasars.\end{abstract}


\section{Introduction} \label{sec:1}
The X-ray emission of luminous quasars is widely
interpreted as inverse-Compton radiation produced in a
hot corona above the accretion disk \citep{Sunyaev1980,Yuan2014}. The tight empirical
relation between the X-ray and ultraviolet (UV)
luminosities, commonly parameterized by the
$\alpha_{\rm ox}$--$L_{2500\,\mathring{A}}$ correlation,
suggests a close coupling between the accretion disk
and the X-ray--emitting corona
\citep{Tananbaum1979,Steffen2006,Lusso2016,Chira2026}.
This relation holds over a wide range of luminosities
and provides an important observational constraint on
disk--corona physics.

However, a small fraction of type 1 quasars significantly deviate from this standard relation, exhibiting X-ray emission weaker by factors of $\gtrsim 10$--100 than expected from their UV luminosities. These X-ray--weak quasars challenge the standard picture of disk--corona coupling and have been the subject of extensive investigation. The prototype of this class, PHL~1811, historically defined the phenomenological properties of X-ray--weak quasars, and is X-ray weak by a factor of $\sim$30--100 relative to expectations from the $\alpha_{\rm OX}$--$L_{\text{2500\AA}}$ relation \citep{Leighly2007}. Thereafter, X-ray--weak AGNs have been reported in both local Seyfert galaxies and high-redshift quasars (i.e., \citet{Miniutti2012}, \citet{Wu2012}, \citet{Luo2015}, \citet{Yang2022}, \citet{Liu2022}, \citet{jin2023extreme}, \citet{Huang2023}).

Intrinsically X-ray--weak quasars have traditionally been suggested to be rare. \citet{Gibson2008} find that the fraction of X-ray--weak, non-BAL QSO sources underluminous by a factor of 10 comprises $\leq$2\% of optically selected SDSS QSOs. A broader investigation by \citet{Pu2020} demonstrates the existence of a population of non-BAL X-ray--weak quasars, reporting that the fractions of quasars in the general population that are X-ray weak by factors of $\geq 6$ and $\geq 10$ are $5.8\% \pm 0.7\%$ and $2.7\% \pm 0.5\%$, respectively. On the other hand, significantly larger fractions of intrinsic X-ray weakness have been observed in specific extreme quasar subpopulations. For instance, recent work by \citet{Nardini2019} finds a large fraction ($\sim$25\%) of X-ray--weak quasars in a sample of luminous blue radio-quiet, non-BAL quasars at $z \sim 3$, with no clear evidence of absorption. In addition, \citet{Laurenti2022} report that $\sim$30\% of
intermediate-redshift, highly accreting
($\lambda_{\rm Edd} \gtrsim 1$) quasars are intrinsically X-ray weak
by factors of $\sim$10--80. This result further suggests that extreme
accretion conditions, particularly in highly accreting AGNs and NLS1-like
systems, may be closely connected to the production of X-ray--weak
states.

Two broad classes of physical explanations have been proposed for X-ray--weak quasars. In intrinsic scenarios, the corona is physically weak/inefficient or disk-reflection dominated, leading to suppressed X-ray emission (e.g., \citealt{Leighly2007,Miniutti2012}). Alternatively, in absorption-driven models, the central engine remains intrinsically normal, but the observed X-ray emission is attenuated by high-column-density, often ionized gas along the line of sight, such as shielding material associated with disk winds \citep[e.g.,][]{Wu2011,Luo2015,Ni2018}.

X-ray--weak quasars are frequently associated with
weak-line quasars (WLQs), a population characterized
by unusually weak high-ionization emission lines\citep[e.g.,][]{Fan1999,Diamond-Stanic2009,Plotkin2010b,Cheng2025, Reich2026}.
Previous studies have shown that approximately 30\%
of X-ray--weak quasars exhibit weak UV emission lines \citep{Wu2011}, while roughly half of WLQs display X-ray weakness
(e.g., \citealt{Luo2015}). This connection suggests
that the physical conditions responsible for weak
line emission, such as shielding gas or modified
ionizing continua, may also play a role in regulating
the observed X-ray emission.

Despite such significant progress, distinguishing between
intrinsic and absorption-driven scenarios remains
observationally challenging, particularly for those objects
identified in persistently X-ray--weak states. Multi-wavelength variability provides a critical
diagnostic. Almost all of the reported X-ray weak quasars exhibits dramatic X-ray
variability with comparatively modest UV and optical
changes \citep[e.g.,][]{Miniutti2012, Ni2020, Ni2022, Liu2022, Huang2026}.
Strong wavelength dependent variability, especially
when accompanied by stable optical emission, indicates
that the variability is likely confined to the compact
X-ray–emitting region or to material along the line
of sight, rather than reflecting a global change in
the accretion flow. In this context, quasars that
transition between X-ray--normal and X-ray--weak
states offer a valuable opportunity to probe the
underlying physical mechanisms. 

Extreme X-ray variations with amplitude exceeding a factor of 10 or more in AGNs are still rare (e.g.,  \citealt{gibson2012x}; \citealt{middei2017long}; \citealt{timlin2020frequency}; \citealt{Wang2025}). Strong X-ray variability events have been observed in a few typical type 1 AGNs (e.g. \citealt{wang2022transienttransient}; \citealt{mehdipour2021transient}; ), which were attributed to changes in the column density of the dust-free obscuring material along the observer's line of sight. High amplitude changes in X-ray flux were also found in a few cases of changing-look AGNs, associated with significant X-ray spectral variability (e.g., \citealt{krumpe2017close};  \citealt{jana2021broad}; \citealt{grupe2015ic}; \citealt{ai2020x}; \citealt{liu2020large}; \citealt{yang2023probing}). This extreme X-ray variation can be interpreted in the scheme of destruction and re-creation of the inner accretion disc and corona. 

Strong and sometimes rapid X-ray variability has been observed in some narrow-line Seyfert 1 galaxies (NLS1s), which are considered to have high or even super-Eddington accretion rates (e.g.,  \citealt{Ai2011}; \citealt{reeves2019momentum}; \citealt{boller2021extreme}; \citealt{2021MNRAS.508.1798P}; \citealt{jin2023extreme}). Several physical mechanisms have been
proposed to explain such extreme X-ray variability in highly accreting
AGNs, including intrinsic changes in the X-ray corona,
variable absorption by clumpy circumnuclear material, and radiatively
driven accretion-disk winds. In particular, rapid changes in coronal
geometry or energetics may strongly modulate the X-ray emission while
producing only limited optical/UV variability
(e.g., \citealt{Miniutti2012, Ricci2020, Wu2020}). At the same time,
powerful disk winds launched via radiation pressure are generally
expected in highly accreting systems
(e.g., \citealt{Giustini2019}; \citealt{jiang2019global};
\citealt{yang2024wind}), and may both intermittently obscure the
compact X-ray source and regulate the disk--corona structure.

In this paper, we report the discovery of a dramatic X-ray state transition observed in a radio-quiet quasar SDSS~J000532.84+200717.4 ($z = 0.3814$, hereafter SDSS~J0005+2007), which undergoes a pronounced transition from an X-ray--normal state to an X-ray--weak phase. SDSS~J0005+2007 classified as a Narrow-line Seyfert 1 galaxy in \citet{Paliya2024}, and it appears as an unresolved source in optical. By combining multi-epoch X-ray spectroscopy, optical
spectroscopy, and long-term multi-wavelength
photometric monitoring, we place constraints on the
physical origin of the X-ray transition. 
The X-ray observations and data reduction are presented in Section~2. In Section~3, we analyze the X-ray spectra, variability and present the multi-band variability. We then discuss the physical interpretation of the extreme X-ray variability in Section~4. Finally we provide a summary in Section~5.
Throughout this paper, we adopt a flat cosmology with
$H_0 = 70\ \rm km\ s^{-1}\ Mpc^{-1}$,
$\Omega_{\rm M} = 0.3$, and
$\Omega_{\Lambda} = 0.7$.

\section{Observations and data reduction} \label{sec:2}

\subsection{X-ray observations and data reduction}\label{sec:21}
SDSS~J0005+2007 was serendipitously detected in the field of view of 14 individual \textit{XMM-Newton} observations \citep{Jansen2001}, as shown in Table~\ref{tab:obslog}. All observations were taken in full-frame mode, except for those obtained in 2000 and 2006, which were performed in partial-window mode. Data from some
individual EPIC cameras were excluded because the source fell near CCD edges or gaps, outside the field of view, or on defective CCD columns. 

The European Photon Imaging Camera (EPIC) data were processed using the Science Analysis System (SAS v21.0.0) and the calibration files generated in April 2024. We reduced the EPIC data following the standard procedure described in the SAS Data Analysis Threads. For the EPIC, only single and double events were selected, and bad pixels and high background flares were filtered from the calibrated event lists based on the standard selection criteria. Source spectra were extracted from circles regions with radii ranging from $20\arcsec$ to $30\arcsec$; background spectra were extracted from source-free regions with a radius of at least 3 times the radius of source regions on the same CCD chip.

We define high- and low-flux states using MOS count-rate threshold of
$5\times10^{-3}\ {\rm counts\ s^{-1}}$ in the
0.2--10~keV band. Observations above this value are classified as
high-flux states, while those below it are classified as low-flux states. For the purposes of subsequent analysis, we designate the first five
epochs, during which the source was relatively bright, as H1--H5, and
the remaining epochs, during which the source was significantly
fainter, as L1--L9. Among the nine low-state observations, the source
was undetected in L3, L5, L6, and L7. For these epochs, we derived
$3\sigma$ upper limits on the source count rates using the
\texttt{uplimit} task within the \texttt{ximage} package, adopting a
circular extraction region with a radius of $60^{\prime\prime}$ centered on the optical position of the source. The results are shown in Table~\ref{tab:obslog}.

\begin{figure}
 \begin{center}
  \includegraphics[width=8cm]{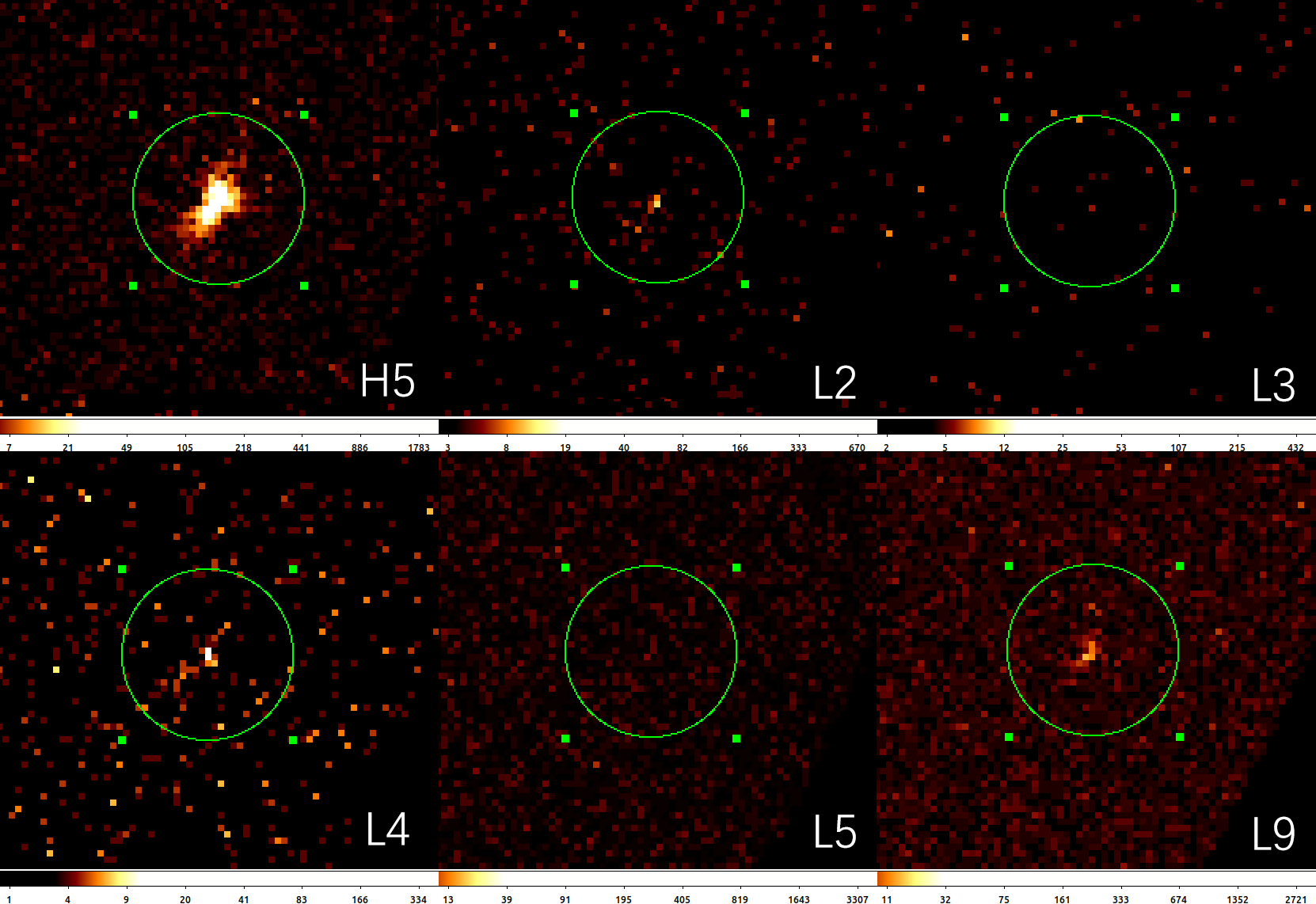}
 \end{center}
\caption{Representative \textit{XMM-Newton} EPIC images of SDSS~J0005+2007 at
different flux states. The source is clearly detected in the high
state (H5), while it becomes weak or undetectable in several low
states (L2, L3, L5), demonstrating extreme long-term X-ray variability. 
The green circle in each image is the circular region of 54$\arcsec$ with a radius centered on the SDSS position of the source.}
\label{fig:image}
\end{figure}

Figure~\ref{fig:image} shows representative \textit{XMM-Newton}/EPIC images of SDSS~J0005+2007 in the 0.2--10~keV band at different flux states. During the high state (e.g., H5), the source is clearly detected with a
compact, high-significance X-ray counterpart. In contrast, during
several low-state observations (e.g., L2, L3, and L5), the source is
either marginally detected or undetected, with the counts
within the extraction region consistent with the background level.
Other low-state epochs (e.g., L4 and L9) show weak but significant
detections. These images provide a clear view of the dramatic X-ray variability of SDSS~J0005+2007.

\begin{figure}
 \begin{center}
  \includegraphics[width=8cm]{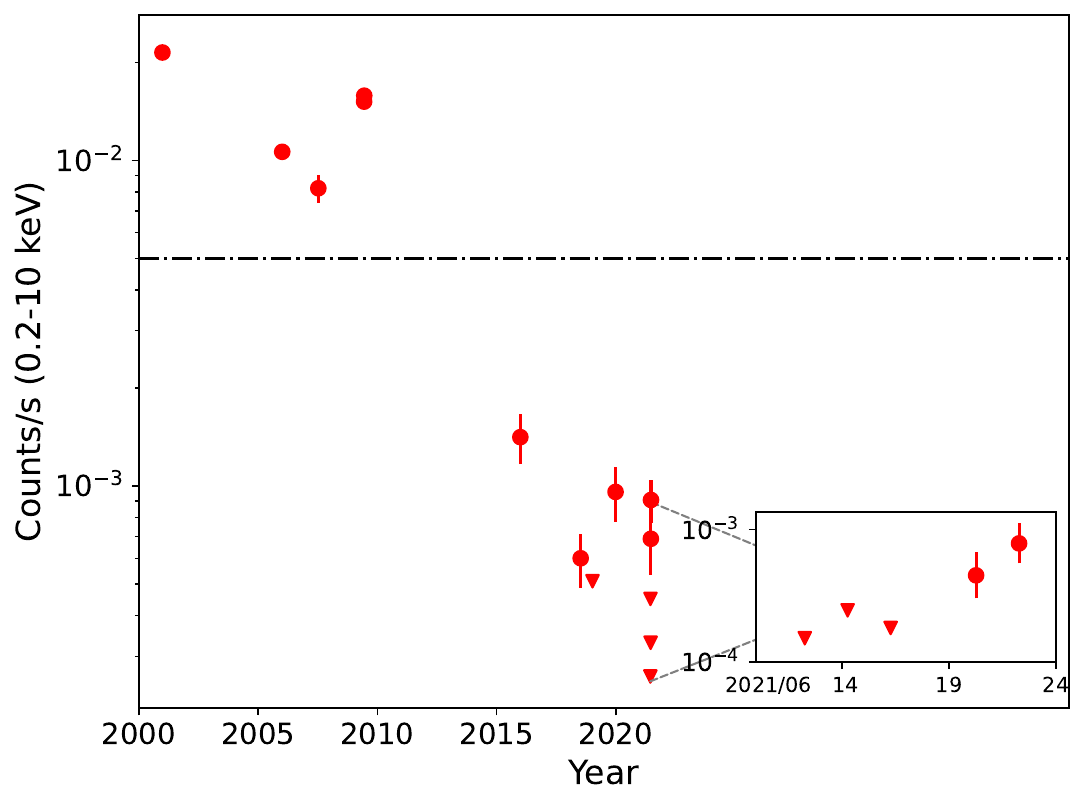} 
 \end{center}
\caption{The \textit{XMM-Newton} EPIC-MOS light curve of SDSS~J0005+2007 in the 0.2--10~keV band. The horizontal dot-dashed line marks the adopted count-rate
threshold of $5 \times 10^{-3}$ cts~s$^{-1}$, which separates the
X-ray high- and low-flux states. The inset shows closely spaced observations in 2021, revealing significant short-term variability with flux changes of a factor of a few on day timescales. Downward arrows indicate $3\sigma$ upper limits.}\label{fig:catarate}
\end{figure}

The \textit{XMM-Newton} X-ray light curve is displayed in Figure~\ref{fig:catarate}. 
The source exhibits pronounced variability over nearly two decades, with count rates decreasing by more than an order of magnitude from the high states observed between 2000 and 2010 to the low states detected after 2015. In particular, the source entered a prolonged X-ray--weak state during 2015--2021, with count rates of $\sim(2$--$10)\times10^{-4}~\mathrm{cts~s^{-1}}$,
compared to $\sim10^{-2}~\mathrm{cts~s^{-1}}$ in earlier epochs. In the sequence of serendipitous \textit{XMM-Newton} observations obtained in 2021 (inset), the source remained X-ray weak but exhibited significant short-term
variability, including a gradual brightening by a factor of $\sim3$ within $\sim10$~days. Several observations yield only upper limits, indicating that the source likely fluctuates around the detection threshold during this low state. Overall, the variability amplitude exceeds a factor of $\sim50$ between the historical high and recent low states, demonstrating extreme long-term X-ray variability.

To investigate the long-term X-ray variability of
SDSS~J0005+2007, we compiled multi-epoch X-ray data
from several missions. In addition to the
\textit{XMM-Newton} observations, we utilized archival
measurements from the Second \textit{Swift} X-ray
Telescope Point Source Catalog (2SXPS; \citealt{Evans2020}) and the
\textit{ROSAT} All-Sky Survey Faint Source Catalog
(RASS-FSC; \citealt{Voges2000}). We also include a
recent observation obtained with the Einstein Probe
Follow-up X-ray Telescope (EP-FXT; \citealt{Yuan2022}); the data reduction procedure for this
observation is described in Section \ref{sec:32}.

\begin{deluxetable*}{ccccccc}
\tabletypesize{\scriptsize}
\tablewidth{0pt}
\tablecaption{X-ray Observation log of SDSS~J0005+2007} \label{tab:obslog}  
\tablehead{\colhead{Obs.ID} & \colhead{Obs.date} & \colhead{Exposure} & \colhead{Instrument} & \colhead{cts s$^{-1}$ (10$^{-3}$)} & \colhead{Off-axis$^{b}$} & \colhead{Abbr.} }
\startdata
& (yyyy-mm-dd) & (ks) &  & (0.2--10keV) & (arcmin) & \\
0101040101& 2000-12-25 & 33.8, 33.8 & MOS1, MOS2 & 17.1 $\pm$ 0.77, 18.10 $\pm$ 0.78 & 12.7 & H1 \\
0306870101& 2006-01-03 & 117.6, 117.6 & MOS1, MOS2 & 12.08 $\pm$ 0.34, 10.63 $\pm$ 0.32 & 12.6 & H2 \\
0510010701& 2007-07-10 & 15.5, 20.0, 20.1 & pn, MOS1, MOS2  & 3.67 $\pm$ 0.18, 7.36 $\pm$ 0.69, 6.93 $\pm$ 0.67 & 11.6 & H3 \\
0600540601& 2009-06-11 & 114.7, 116.1 & MOS1, MOS2  & 15.31 $\pm$ 0.41, 15.16 $\pm$ 0.43 & 11.6 & H4 \\
0600540501& 2009-06-13 & 80.9, 80.9 & MOS1, MOS2 & 16.94 $\pm$ 0.49, 15.81 $\pm$ 0.48 & 11.6 & H5 \\
0741280201& 2015-12-30 & 69.3 & MOS2 & 1.41 $\pm$ 0.25 & 12.6 & L1 \\
0780500301& 2018-07-11 & 107.0 & MOS2 & 0.60 $\pm$ 0.11 & 11.6 & L2 \\
0831790601& 2019-01-08 & 89.2 & MOS2  & 0.51$^{a}$ & 12.6 & L3 \\
0854590401& 2019-12-27 & 98.6 & MOS2 & 0.96 $\pm$ 0.18 & 12.5 &  L4 \\
0842761401& 2021-06-14 & 85.3 & MOS2 & 0.26$^{a}$ & 12.6 & L5 \\
0842760201& 2021-06-16 & 87.2 & MOS2 & 0.45$^{a}$ & 11.8 & L6 \\
0842761101& 2021-06-18 & 86.7 & MOS2 & 0.33$^{a}$ & 11.7 & L7 \\
0842761201& 2021-06-20 & 89.1, 89.7 & MOS1, MOS2 &0.66 $\pm$ 0.17, 0.69 $\pm$ 0.16 & 11.7 & L8 \\
0842761301& 2021-06-22 & 88.3, 88.4 & MOS1, MOS2 &1.10 $\pm$ 0.15, 0.91 $\pm$ 0.14 & 11.7 & L9 \\
\enddata
\tablenotetext{a}{The 3$\sigma$ upper limits at the epochs of L3, L5, L6, L7. \vspace{-1.5ex}}
\tablenotetext{b}{Only the off-axis values of the MOS2 camera are listed here. }
\end{deluxetable*}

\subsection{Optical Spectroscopy}\label{sec:22}

SDSS~J0005+2007 was observed by Sloan Digital Sky Survey (SDSS) on 2014 October 7 through a 3$\arcsec$ diameter fiber over the wavelength range 2600–7500~\AA~at a spectral resolution of $\simeq$2000. Follow-up optical spectroscopy of SDSS~J0005+2007 was conducted using the Double Spectrograph (DBSP; \citealt{Oke1982}) mounted on the Hale 200-inch Telescope at Palomar Observatory (P200) on 2024 September 9. The observations were carried out through a $1.5\arcsec$ slit using a D55 dichroic, which splits the incoming light between the blue arm (600/4000 grating) and the red arm (316/7500 grating). The grating angles were adjusted to provide a nearly continuous wavelength coverage from $\sim 3400$ to $10,000$ \AA\ (corresponding to $\sim 2500$--$7200$ \AA\ in the rest frame). We obtained one more epoch of optical spectroscopy on 2025 December 14 using the Next Generation Palomar Spectrograph (NGPS; \citealt{Jiang2018}) mounted on the P200 with slit width of $1.5\arcsec$. NGPS records simultaneous spectra in two channels, covering approximately 580--780\,nm (R channel) and 760--1040\,nm (I channel). The data reduction was performed using the \texttt{PypeIt} spectroscopic reduction pipeline \citep{Prochaska2020}, following standard long-slit procedures, including bias subtraction, flat-fielding, and wavelength calibration. The spectrum was extracted using an optimal aperture and flux-calibrated using a standard star observed on the same night.

\section{Results}  \label{sec:3}
\subsection{X-ray spectral Analysis} \label{sec:31}
X-ray spectral fitting was performed using \textsc{xspec} v12.13.1 \citep{arnaud}. 
For the high-flux epochs (H1--H5), the spectra were grouped to a minimum of 10 counts per bin. To improve the signal-to-noise
ratio in the faint phase, we constructed a stacked spectrum by combining all low-state observations. Owing to the limited photon statistics, the \textsc{W}-statistic was adopted throughout the analysis. For each epoch, the EPIC pn and/or MOS1/MOS2 spectra were fitted jointly when available.

We first modeled all spectra using a simple power-law modified by Galactic absorption fixed at
$N_{\rm H,Gal} = 1.7 \times 10^{20}~{\rm cm^{-2}}$ \citep{kalberla}. This model provides statistically acceptable fits for most observations. During the high-flux states, the inferred photon indices lie in the range $\Gamma = 2.85$--3.09, whereas the stacked low-state spectrum yields a photon index of $\Gamma \sim 2.78$, indicating a slightly flatter spectral shape.

\begin{figure}
 \begin{center}
  \includegraphics[width=8cm]{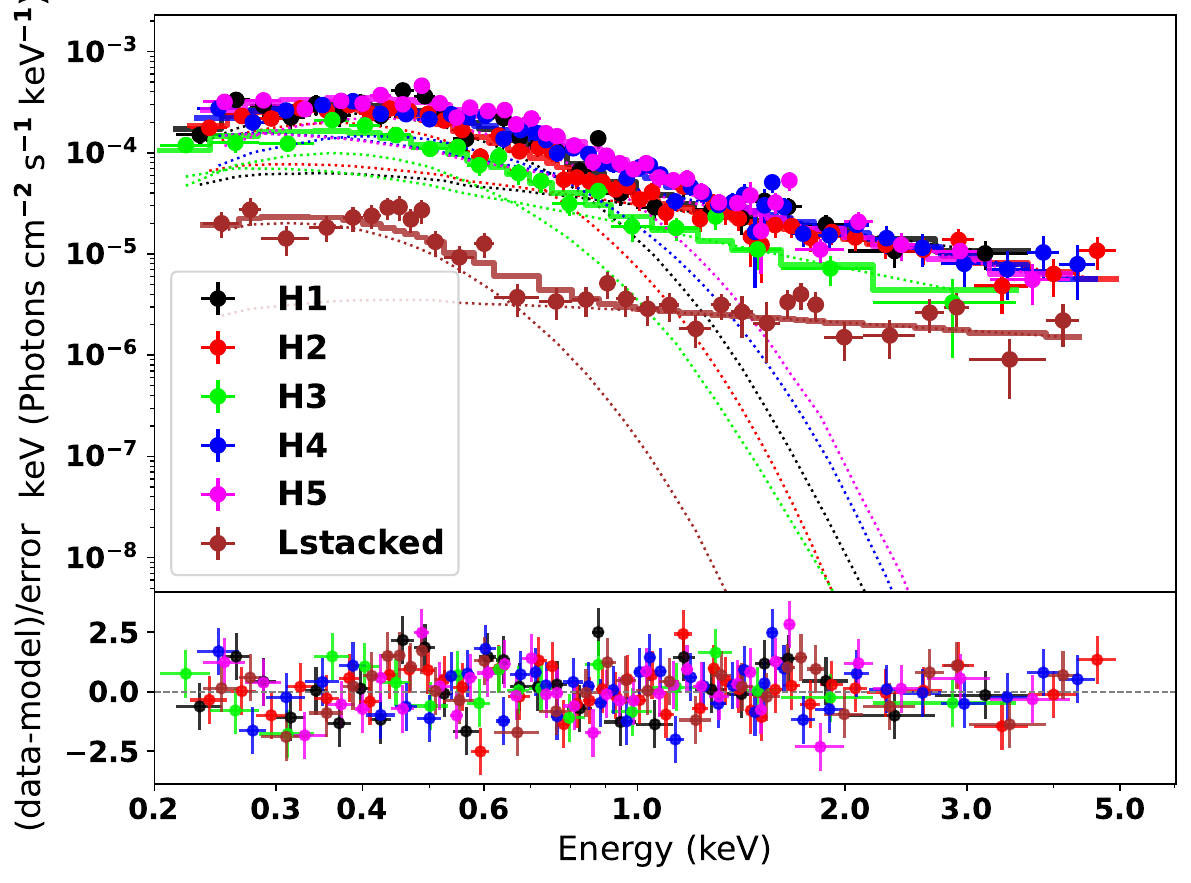}
 \end{center}
\caption{Unfolded \textit{XMM-Newton} EPIC X-ray spectra of SDSS~J0005+2007 in different flux states, shown together with the best-fit spectral models consisting of a Galactic-absorbed power law plus a blackbody component. For clarity, only the EPIC-MOS2 data are displayed. The lower panel shows the residuals.}
\label{fig:alleude}
\end{figure}

We then adopted a composite model consisting of a Galactic-absorbed power law plus a 
blackbody component (\texttt{tbabs*(zpo+zbb)}). 
For the high-flux states, the addition of the blackbody component leads to a significant 
improvement in the fit statistics, with typical C-stat $\sim$ 30 compared to the 
single power-law model. 
The resulting fits are statistically acceptable, as 
illustrated in Figure~\ref{fig:alleude}. 
The best-fit parameters for all epochs are summarized in Table~\ref{tab:result}.

For the stacked low-flux spectrum, the same composite model also
provides an acceptable fit, but with noticeably different best-fit
parameters. In particular, the power-law component becomes
significantly harder ($\Gamma = 1.45^{+0.38}_{-0.37}$) compared to
the high-flux states, while the overall flux is reduced by more than
an order of magnitude. The
addition of an intrinsic absorption component (e.g., \texttt{zphabs})
does not lead to a statistically significant improvement in the fit
for either the high- or low-flux spectra.

The apparent flattening of the X-ray spectrum in the low state may be attributed to partial-covering absorption. To explore this possibility, we fitted the stacked low-state spectrum with an ionized partial-covering absorber model, \textit{zxipcf} \citep{Reeves2008}, while fixing the intrinsic continuum shape (photon index and blackbody temperature) to the high-state (H1) values. This model provides a statistically acceptable fit (C-stat$/\mathrm{d.o.f.}$ = 29.30/28). The best-fit parameters imply an absorber column density of $N_{\rm H} \approx 6.8 \times 10^{22}\ \rm cm^{-2}$, an ionization parameter of $\log (\xi/erg/s/cm) \approx 2.3$, and a high covering fraction ($f_{\rm cov} \approx 0.73$--1.0). However, given the limited photon statistics of the low-state spectrum, these parameters remain poorly constrained.

\begin{deluxetable*}{cccccccc}
\tabletypesize{\scriptsize}
\tablewidth{0pt}
\caption{Parameters of X-ray Spectral Fits Using
\texttt{tbabs*(zpo+zbb)}}
\label{tab:result}

\tablehead{
\colhead{Abbr.} &
\colhead{$T_{\rm b}$} &
\colhead{$\log F_{\rm b}$} &
\colhead{$\Gamma$} &
\colhead{$\log F_{\rm p}$} &
\colhead{$\log F_{\rm soft}^{a}$} &
\colhead{$\log F_{\rm hard}^{b}$} &
\colhead{C-stat/d.o.f.}}
\startdata
H1 &
$0.14_{-0.01}^{+0.01}$ &
$-12.70_{-0.13}^{+0.08}$ &
$2.24_{-0.38}^{+0.36}$ &
$-12.88_{-0.07}^{+0.07}$ &
$-12.38_{-0.04}^{+0.03}$ &
$-12.73_{-0.05}^{+0.04}$ &
46.21/29 \\
H2 &
$0.14_{-0.01}^{+0.01}$ &
$-12.88_{-0.09}^{+0.06}$ &
$2.26_{-0.21}^{+0.21}$ &
$-12.93_{-0.04}^{+0.04}$ &
$-12.60_{-0.02}^{+0.02}$ &
$-12.78_{-0.03}^{+0.02}$ &
55.29/44 \\
H3 &
$0.13_{-0.02}^{+0.01}$ &
$-13.03_{-0.27}^{+0.12}$ &
$2.25_{-0.54}^{+0.48}$ &
$-13.05_{-0.07}^{+0.06}$ &
$-12.74_{-0.04}^{+0.04}$ &
$-12.90_{-0.05}^{+0.05}$ &
3.66/7 \\
H4 &
$0.18_{-0.01}^{+0.01}$ &
$-12.88_{-0.12}^{+0.10}$ &
$2.54_{-0.19}^{+0.16}$ &
$-12.63_{-0.03}^{+0.03}$ &
$-12.38_{-0.02}^{+0.02}$ &
$-12.54_{-0.02}^{+0.02}$ &
42.20/43 \\
H5 &
$0.17_{-0.01}^{+0.01}$ &
$-12.65_{-0.16}^{+0.11}$ &
$2.46_{-0.43}^{+0.35}$ &
$-12.69_{-0.07}^{+0.06}$ &
$-12.26_{-0.03}^{+0.03}$ &
$-12.58_{-0.04}^{+0.04}$ &
45.86/38 \\
Lstacked &
$0.11_{-0.02}^{+0.02}$ &
$-13.82_{-0.12}^{+0.09}$ &
$1.45_{-0.37}^{+0.38}$ &
$-14.05_{-0.10}^{+0.08}$ &
$-13.66_{-0.07}^{+0.05}$ &
$-13.58_{-0.10}^{+0.10}$ &
41.28/33 \\
\enddata
\tablenotetext{}{Notes: Columns 2--7 list the blackbody
temperature in keV ($T_{\rm b}$), blackbody flux in 0.2--2~keV ($F_{\rm b}$),
photon index ($\Gamma$), power-law flux in 0.2--2~keV ($F_{\rm p}$),
total soft-band flux in 0.2--2~keV ($F_{\rm soft}$), and hard-band flux
in 2--10~keV ($F_{\rm hard}$), respectively. All fluxes are given in
units of erg s$^{-1}$ cm$^{-2}$.}
\end{deluxetable*}

\subsection{X-ray and multi-wavelength Variability}\label{sec:32}
The X-ray emission of SDSS~J0005+2007 shows extreme variability over both long
and short timescales within the \textit{XMM-Newton} epochs, as shown in Figure~\ref{fig:catarate}. To extend the temporal baseline beyond the \textit{XMM-Newton} era, we
compiled archival X-ray measurements from the ROSAT All-Sky Survey Faint
Source Catalog (RASS-FSC) and the \textit{Swift}/XRT 2SXPS catalog.
These data provide crucial constraints on the long-term evolution of
SDSS~J0005+2007 over more than three decades.

SDSS~J0005+2007 was detected by ROSAT during the all-sky survey with a count
rate of $0.032 \pm 0.013$~cts~s$^{-1}$ in a 261~s exposure.
The reported ROSAT hardness ratio is HR $\approx -1$, indicating that the
source exhibited a very soft X-ray spectrum at this early epoch.
Assuming a Galactic-absorbed power-law model with $\Gamma = 3$, motivated by the soft X-ray spectra
measured with \textit{XMM-Newton} (Section~\ref{sec:31}), this corresponds to an unabsorbed 0.2--2~keV flux of
$4.33 \pm 1.80\times10^{-13}$~erg~cm$^{-2}$~s$^{-1}$.
This result demonstrates that SDSS~J0005+2007 was X-ray bright and spectrally soft at ROSAT epoch.

\begin{figure}
 \begin{center}
  \includegraphics[width=8cm]{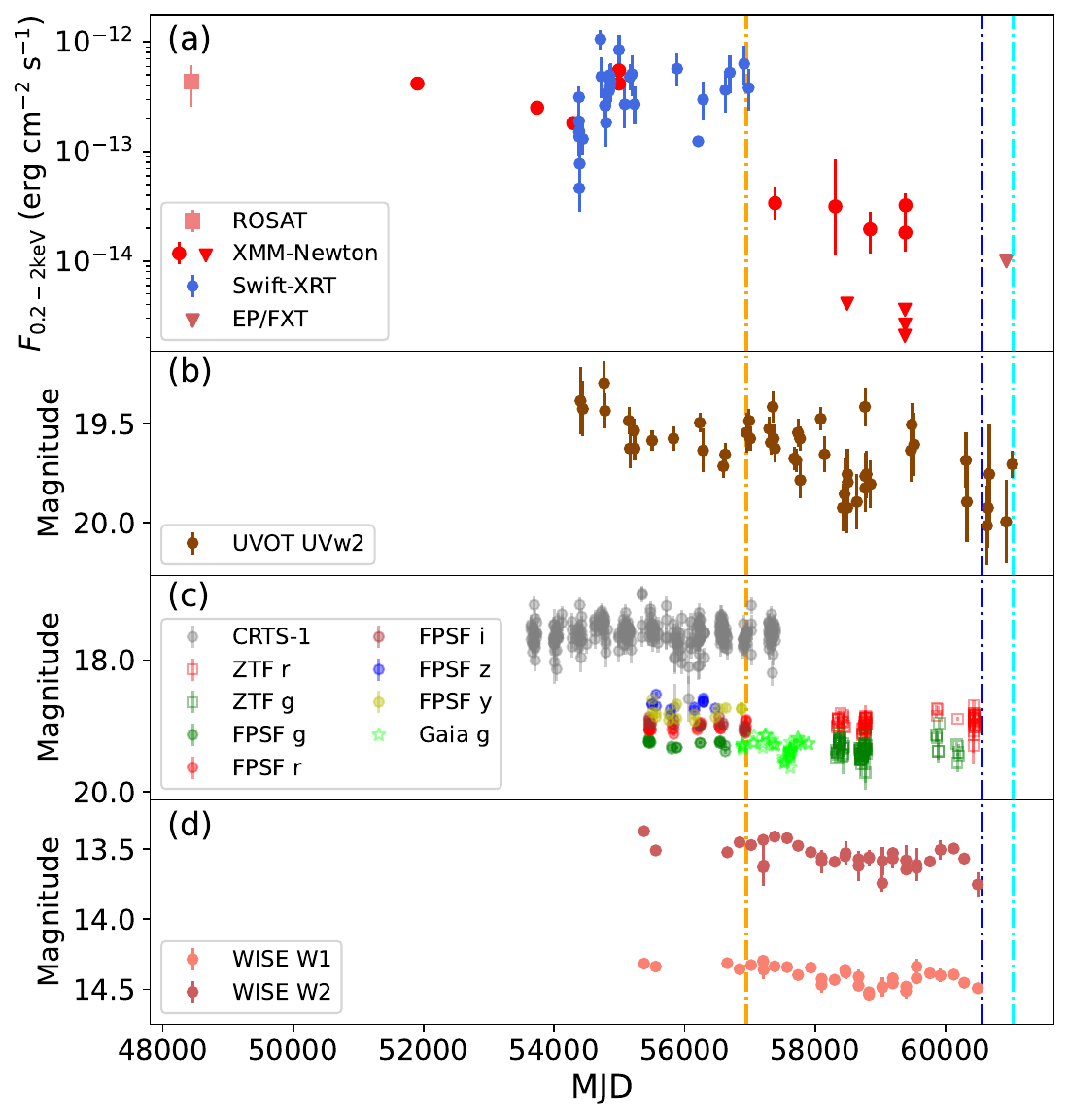} 
 \end{center}
\caption{Multi-wavelength light curves between 1991--2025 of SDSS~J0005+2007. (a) X-ray light curve in the 0.2--2~keV band compiled from \textit{XMM-Newton}, ROSAT, \textit{Swift}/XRT, and EP-FXT observations. For epochs with non-detections, 3$\sigma$ upper limits are shown. The vertical dot-dashed lines mark the epochs of the SDSS and P200 spectroscopic observations. (b) Ultraviolet light curve from \textit{Swift}/UVOT in the UVW2 band. (c) Optical light curves from CRTS, ZTF, Pan-STARRS, and \textit{Gaia}. (d) Mid-infrared light curves from the \textit{WISE}/NEOWISE W1 and W2 bands, which were binned into 7-day intervals with the mean magnitude shown.
}\label{fig:al_lc}
\end{figure}

There are multiple epochs of X-ray detections of SDSS~J0005+2007 in the \textit{Swift}/XRT 2SXPS catalog.
The \textit{Swift}/XRT count rates were converted to unabsorbed fluxes assuming the same
Galactic-absorbed power-law model with $\Gamma = 3$.
As shown in Figure~\ref{fig:al_lc}(a), the \textit{Swift} fluxes are consistent with the historical high-flux level
observed during 2000--2010, prior to the transition into the prolonged X-ray--weak state after 2015.

A recent ToO observation with the Einstein Probe Follow-up X-ray Telescope
(EP-FXT) provides an independent constraint on the current X-ray state
of SDSS~J0005+2007. The source was not detected in the EP-FXT observation obtained on
2025 September 15.
The resulting 3$\sigma$ upper limit on the count rate is
$1.67\times10^{-4}$~cts~s$^{-1}$ in the 0.5--10~keV band.
Assuming the same Galactic-absorbed power-law model ($\Gamma = 3$),
this corresponds to a conservative unabsorbed flux upper limit of
$\lesssim1\times10^{-14}$~erg~cm$^{-2}$~s$^{-1}$.
This non-detection confirms that SDSS~J0005+2007 remains in a deep
X-ray low-flux state at the most recent epoch (Figure~\ref{fig:al_lc}(a)).

To investigate the rest-frame ultraviolet behavior of SDSS~J0005+2007,
we utilized photometric data from the Ultraviolet/Optical Telescope
(UVOT; \citealt{Roming2005}) onboard the \textit{Swift} observatory.
The observations were primarily obtained with the UVW2 filter
($\lambda_{\rm eff}=1928$~\AA), which probes the UV emission near rest-frame 1400~\AA for this quasar. We retrieved the level-2 image files from the HEASoft archive and performed aperture photometry using the \texttt{uvotsource} task.
Source counts were extracted from a circular region with a radius of
$7\arcsec$ centered on the optical position of the quasar.
This aperture, slightly larger than the standard $5\arcsec$ radius,
was adopted to account for the extended point-spread function and to
ensure full flux enclosure. The background was estimated from a concentric, source-free annular region. The measured magnitudes were corrected for Galactic extinction using the the extinction law from \citet{Schlafly2011}.

We show the UVW2 light curve in Figure~\ref{fig:al_lc}b. It is clear that the UV emission of SDSS~J0005+2007 shows a mild long-term fading trend that broadly
follows the evolution of the X-ray emission, but with a substantially smaller amplitude of $\sim0.3$~mag. In contrast to the dramatic X-ray decline, no sharp UV drop is observed. Instead, the UV flux varies gradually, with epochs of strong X-ray suppression or non-detection generally corresponding to relatively fainter UV levels. 

To investigate the optical/mid-infrared variability of SDSS~J0005+2007, we compiled archival optical and
mid-infrared photometry from Catalina Real-Time Transient Survey (CRTS; \citealt{drake}), Zwicky Transient Facility (ZTF; \citealt{Masci2019}), Panoramic Survey Telescope and Rapid Response System (Pan-STARRS; \citealt{Flewelling2020}), \textit{Gaia} \citep{Gaia2022}, and Near-Earth Object Wide-field Infrared Survey Explorer Reactivation Mission (\textit{NEOWISE}; \citealt{Mainzer2011, WISE2020b}). We used the forced photometry from ZTF and Pan-STARRS. 
As shown in Figure~\ref{fig:al_lc}, the optical and mid-infrared light curves do not exhibit
any substantial long-term variability, remaining stable over timescales of years to decades.
The observed variability amplitudes are much smaller than those seen in X-rays,
indicating a clear decoupling between the extreme X-ray variability and the emission at longer wavelengths.

\subsection{Optical Spectroscopy}\label{sec:33}

\begin{figure*}
 \begin{center}
  \includegraphics[width=18cm]{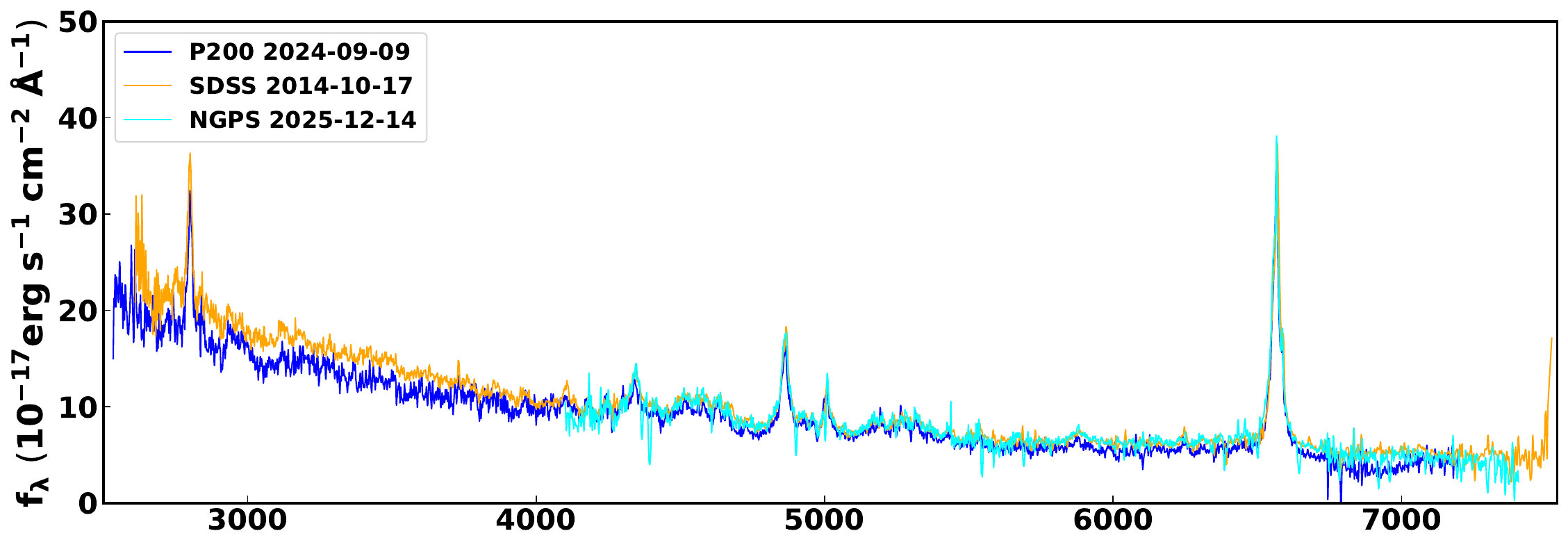} 
 \end{center}
\caption{Rest-frame optical spectra of SDSS~J0005+2007 obtained at three epochs: 2014 October 17 from SDSS (orange line), 2024 September 09 from P200/DBSP (blue line), and 2025 December 14 from P200/NGPS (cyan line). All spectra have been corrected for Galactic extinction and shifted to the rest frame with $z=0.3814$. The fluxes of the P200 and NGPS spectra were recalibrated by scaling their narrow [\ion{O}{3}] $\lambda$5007 emission line flux to match that of the SDSS spectrum.}
\label{fig:PvS}
\end{figure*}

Figure~\ref{fig:PvS} presents the rest-frame optical spectra of SDSS~J0005+2007 obtained at three
different epochs. All spectra have been corrected for Galactic extinction and shifted to the
rest frame assuming a redshift of $z = 0.3814$. The P200/DBSP and P200/NGPS spectra were
flux-recalibrated by scaling their narrow [O~III] $\lambda5007$ emission-line fluxes to match
that of the SDSS spectrum. The optical spectra of SDSS~J0005+2007 remain remarkably stable over timescales of years to decades. Despite the extreme
variability observed in the X-ray band, the broad emission lines have remained remarkably
stable over the past decade. This behavior suggests that the photoionizing continuum incident on the broad-line region has not undergone significant long-term changes or been obscured.

We modeled the SDSS optical spectrum using the \texttt{PyQSOFit}
code \citep{guo} and performed spectral decomposition to derive the
key physical parameters of the source. After subtracting the
host-galaxy contribution, we measured an AGN continuum flux density at
5100~\AA\ of
$F_{5100} = 7.07 \times 10^{-17}\ \mathrm{erg\ s^{-1}\ cm^{-2}\ \AA^{-1}}$
and a full width at half maximum (FWHM) of the broad H$\beta$
component of
$1823.32\ \mathrm{km\ s^{-1}}$. These values are consistent with those
reported by \citet{Rakshit2020}, who also classified the source as an
NLS1 galaxy. Using the continuum luminosity at 5100~\AA\
($L_{5100}$) and the FWHM of the broad H$\beta$ emission line, we
estimated the single-epoch (SE) virial black hole mass following the
calibration of \citet{Shen2024}, obtaining
$M_{\mathrm{BH}} \approx 3.18 \times 10^{7}\ M_{\odot}$ from the
\textit{SDSS} spectrum. Applying the bolometric correction of
\citet{DallaBonta2020}, we derived a bolometric luminosity of
$L_{\mathrm{bol}} \approx 2.56\times10^{45}\ \mathrm{erg\ s^{-1}}$
and an Eddington ratio of
$\lambda_{\mathrm{Edd}} \approx 0.614$. We further estimated the black
hole mass using the empirical relation based on the broad H$\alpha$
luminosity and the FWHM of the broad H$\alpha$ emission line
\citep{Greene2005}, obtaining $M_{\mathrm{BH}} \approx 1.06 \times 10^{7}\ M_{\odot}$.

As discussed by
\citet{DallaBonta2020}, SE virial black hole masses based on broad
H$\beta$ may be systematically overestimated in highly accreting AGNs.
We therefore additionally applied Equation~(30) of
\citet{DallaBonta2020}, which accounts for the Eddington-ratio
dependence, obtaining an alternative estimate of
$M_{\mathrm{BH}} \approx 2.66 \times 10^{6}\ M_{\odot}$. The Eddington-ratio--corrected black hole mass is approximately one order of magnitude lower than the standard SE virial estimate, implying that SDSS~J0005+2007 may be accreting at an even higher
Eddington ratio than inferred from the uncorrected mass estimate.

From the same spectral fits, we measured the rest-frame equivalent widths (REWs) of the \ion{Mg}{2} emission line, finding $\rm REW_{\rm Mg\,II} \approx 9.68$~\AA\ from the SDSS spectrum and $\approx 24.32$~\AA\ from the P200 spectrum. These values are lower than the typical range for quasars ($\sim 30$--$50$~\AA) and fall within the regime commonly associated with WLQs \citep[e.g.,][]{Diamond-Stanic2009,Plotkin2010b}. However, a definitive WLQ classification generally requires constraints on high-ionization lines, most notably $\rm REW_{\rm C\,IV} \lesssim 10$~\AA. In the absence of rest-frame UV spectroscopy, we cannot conclusively determine whether SDSS~J0005+2007 belongs to the WLQ population. Nevertheless, the relatively weak \ion{Mg}{2} emission suggests potential physical similarities, such as the possible presence of shielding gas that modifies the ionizing continuum incident on the broad-line region. 

Given the possible WLQ-like nature of SDSS~J0005+2007 together
with its high-accretion NLS1 properties, previous studies have
suggested that Fe\,II-corrected virial black hole mass estimates may
be more appropriate for such systems
(e.g., \citealt{Du2019, Ha2023}). From the spectral decomposition, we measured an integrated optical Fe II flux of $F_{\rm Fe} \approx 584.02 \times 10^{-17} \rm \, erg \, s^{-1} \, cm^{-2}$ over the $4434\text{--}4684\,\text{\AA}$ rest-frame range and a broad H$\beta$ flux of $F_{{\rm H}\beta}^{\rm br} \approx 367.24 \times 10^{-17} \rm \, erg \, s^{-1} \, cm^{-2}$, yielding a strong optical Fe II emission strength of $R_{\rm FeII} = F_{\rm Fe}/F_{{\rm H}\beta}^{\rm br} \approx 1.59$. Applying the prescription of
\citet{Du2019}, we derived an Fe\,II-corrected black hole mass of
$M_{\mathrm{BH}} \approx 1.50 \times 10^{7}\ M_{\odot}$, broadly
consistent with the standard H$\beta$-based virial mass estimate.

\subsection{Spectra Energy Distribution}\label{sec:34}

\begin{figure}
 \begin{center}
  \includegraphics[width=8cm]{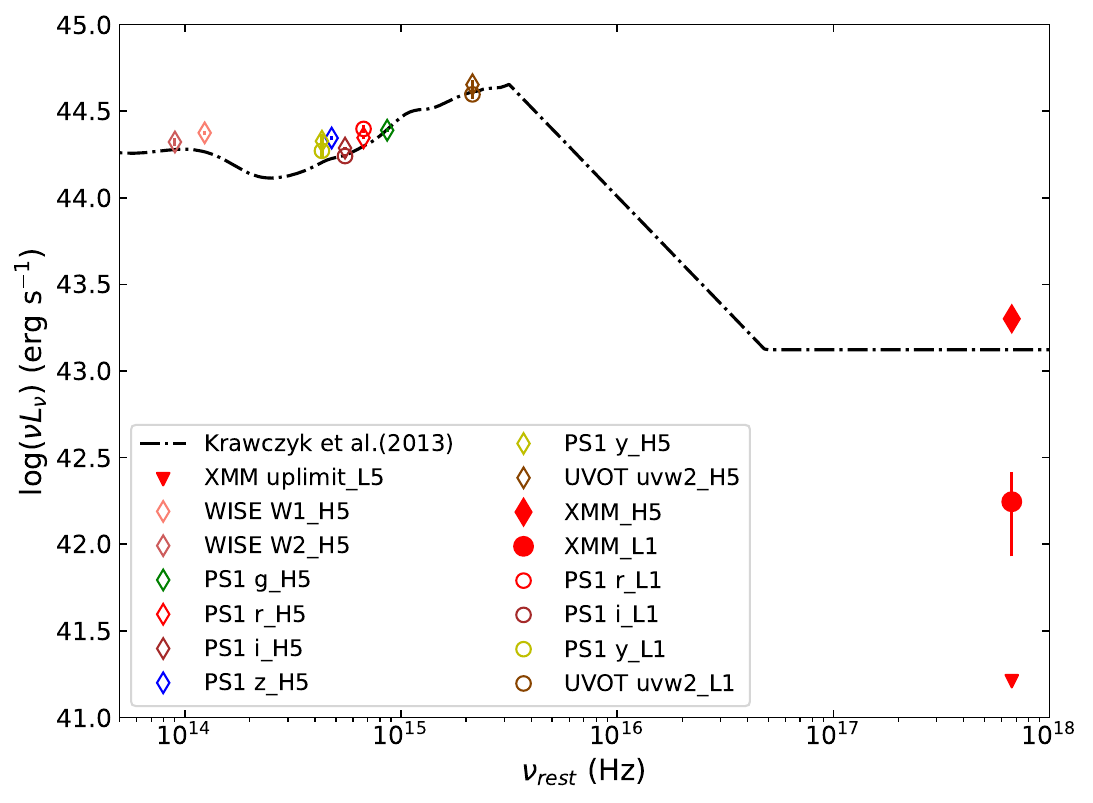} 
 \end{center}
\caption{Infrared-to-X-ray spectral energy distributions (SEDs) of SDSS~J0005+2007 in the X-ray high state and low state. The IR–to–UV data points represent quasi-simultaneous observations during the X-ray high state
(H5; diamonds) and the low state (L1; circles).
The red downward triangle denotes the $3\sigma$ upper limit from the undetected X-ray epoch (L5).
For comparison, the black dot–dashed line shows the mean SED of radio-quiet quasars from \citet{Krawczyk2013}, normalized to the 5100~\AA\ flux of SDSS~J0005+2007.
}\label{fig:sed}
\end{figure}

Figure~\ref{fig:sed} presents the rest-frame broadband SED of SDSS~J0005+2007,
combining multi-wavelength photometry from the infrared to the X-ray bands.
We compare the data obtained during the X-ray high state (H5) and the low
state (L1/L5) with the standard radio-quiet quasar template of
\citet{Krawczyk2013}. In the high state (H5), the source emission (diamonds)
closely follows the standard template across the infrared, optical, UV, and
X-ray bands, with the X-ray luminosity
consistent with that expected for a typical type~1 quasar.

In contrast, the low state reveals a clear decoupling between the optical–UV and X-ray emission components.
In the infrared and optical bands, the
low-state data points (circles) remain comparable to those in the high state, showing only modest scatter without a systematic flux decrease. The UV emission
exhibits a mild decline, but its amplitude remains small compared to the dramatic changes observed in the X-ray regime. The X-ray flux in the low state drops by more than an order of magnitude,
and the deepest upper limit at epoch L5 lies nearly two orders of magnitude
below the template prediction. This SED evolution indicates that the extreme X-ray weakness occurs while the optical–UV emission remains broadly consistent with that of a luminous accretion disk.

To systematically quantify this behavior, we computed the X-ray--to--optical power-law slope,
$\alpha_{\rm ox}=0.3838 \log(L_{\rm 2\,keV}/L_{2500\,\mathring{A}})$
\citep{Tananbaum1979}, for five epochs (H3, H5, L1, L4, and L9) with
quasi-simultaneous UV and X-ray observations. The rest-frame
2500~\AA\ luminosities were derived from the UVW2 photometry,
assuming a spectral slope of $\alpha_{\lambda} = -1.57$, inferred
from the SDSS spectral modeling.
During the high states, the inferred $\alpha_{\rm ox}$ values range from
$-1.33$ to $-1.53$, consistent with those of typical quasars.
In contrast, the low-state measurements (L1, L4, and L9) show significantly
steeper slopes of $\alpha_{\rm ox} \lesssim -1.70$, while the deepest
non-detection at epoch L5 implies an upper limit of
$\alpha_{\rm ox} \approx -2.13$.

To further characterize the X-ray weakness, we computed
$\Delta \alpha_{\rm ox} = \alpha_{\rm ox,obs} - \alpha_{\rm ox,exp}$,
where $\alpha_{\rm ox,exp}$ is the expected value from the
$\alpha_{\rm ox}$--$L_{2500\,\mathring{A}}$ relation of
\citet{Steffen2006}. The low-state observations yield
$\Delta \alpha_{\rm ox} \lesssim -0.3$, corresponding to an X-ray
weakness factor of $f_{\mathrm{weak}} = 10^{-\Delta \alpha_{\mathrm{ox}} / 0.384} \gtrsim 6$, placing the source in the
X-ray--weak regime (e.g., \citealt{Pu2020}).

\section{Discussion}\label{sec:4}
The luminous NLS1 SDSS~J0005+2007 presents extreme X-ray variability over timescales of decades, in particular a systematic decline in the soft X-ray flux by a factor of $\sim10$ over a timescale of $\sim5$~yr. In contrast, the UV emission varies only mildly ($\sim0.3$~mag), while the optical continuum, broad emission-line profiles, and mid-infrared emission remain remarkably stable over decade-long timescales. During the high states, the X-ray spectrum is persistently soft ($\Gamma \gtrsim 2.5$--3), typical of luminous type~1 quasars, whereas the stacked low-state spectrum shows substantial flux suppression together with indications of spectral hardening. These combined variability and spectral properties provide the observational framework for interpreting the physical origin of the X-ray transition.

\subsection{Nature of the X-ray Weakness}

A key question raised by the extreme X-ray variability of
SDSS~J0005+2007 is whether it reflects a global change in the accretion
flow. Several observational results argue against this interpretation.
The optical and mid-infrared emission remain stable over timescales of
years to decades, and multi-epoch optical spectroscopy reveals no
significant long-term changes in either the continuum level or the
broad emission-line profiles. The persistence of the broad-line region
(BLR) properties indicates that the photoionizing continuum incident on
the line-emitting gas has not undergone substantial long-term
variation. In addition, the UV luminosity exhibits only mild
variability compared to the dramatic X-ray evolution. Taken together,
these results suggest that the optical--UV emitting accretion disk
remains largely intact during the X-ray low states, disfavoring a
global shutdown of accretion.

In contrast, the X-ray emission declines by more than an order of
magnitude, and by nearly two orders of magnitude in the deepest
epoch. Two broad classes of mechanisms may account for such behavior:
absorption along the line of sight or intrinsic changes in the X-ray
corona.

The spectral hardening observed in the stacked low-state spectrum is
qualitatively consistent with obscuration. When the intrinsic continuum
shape is fixed to the high-state values, an ionized partial-covering
absorber ($N_{\rm H} \approx 6.8 \times 10^{22}\ \rm cm^{-2}$,
$\log (\xi  \text{/erg cm s$^{-1}$}) \approx 2.3$) provides an acceptable description of the
low-state spectrum. Although the photon statistics do not tightly
constrain these parameters, this model demonstrates that a moderate
column of ionized gas can reproduce both the observed flux suppression
and the spectral shape. Such a scenario is broadly consistent with the
interpretation proposed for other X-ray--weak quasars. For example,
PHL~1811 analogs  and weak-line quasars are often associated with
shielding gas that attenuates the ionizing continuum (\citep{Wu2011,Luo2015,Ni2018,Ni2022}). The presence of disk winds, inferred from weak,
blueshifted, and asymmetric \ion{C}{4} emission lines, and the hard
X-ray spectral properties revealed by stacking analyses further support
an absorption-related origin (\citealp[e.g.][]{Leighly2007, Miniutti2012, Liu2022}) . Recent hard X-ray observations with
\textit{NuSTAR} also indicate that heavy absorption can account for the
apparent X-ray weakness in some cases \citep[e.g.,][]{wang2022transient}.

Alternatively, changes in the X-ray corona itself may also contribute
to the observed variability. Variations in coronal geometry or
radiative efficiency could suppress the X-ray emission while leaving
the accretion disk largely unaffected. At present, the available data
do not uniquely distinguish between complex absorption and intrinsic
coronal variations. Nevertheless, the pronounced wavelength-dependent
behavior --- extreme X-ray variability accompanied by minimal changes
in the optical and UV emission --- indicates that the dominant
mechanism is confined to the compact X-ray--emitting region rather than
reflecting a global transformation of the accretion flow.

\subsection{Short-term Variability: Evidence for Clumpy Structure}\label{sec:42}
Within the prolonged low state, SDSS~J0005+2007 continues to exhibit rapid variability on timescales of days to weeks, including epochs of non-detection followed by rebrightening. Such behavior indicates that the low state is not static. If obscuration dominates, the absorber is unlikely to be homogeneous. The rapid variability instead suggests compact structures moving across the line of sight. A natural interpretation is that the absorbing medium is clumpy, consisting of dense, compact clouds embedded within a more diffuse, partially covering wind. Occasional non-detections can then be explained by individual optically thick clumps transiting the line of sight to the compact X-ray emitting region.

Similar rapid X-ray dimming events have been reported in other quasars. For example, \citet{Liu2022} observed a radio-quiet type~1 quasar at $z \sim 2.6$ whose X-ray flux decreased by a factor of $\sim7.6$ within two rest-frame days, which was interpreted as a fast-moving absorber crossing and fully covering the X-ray emitting corona. Such events support the possibility of compact absorbing structures in quasar winds.

The duration of obscuration events can be used to estimate the physical size of the transiting clumps. The crossing time, $t_{\rm cross}$, for an absorber of projected size $r_{\rm abs}$ moving with Keplerian velocity at an orbital radius $r_{\rm orb}$ is given by
\begin{equation}
t_{\rm cross} \approx 0.7 
\left( \frac{r_{\rm orb}}{\text{light-day}} \right)^{3/2}
M_6^{-1/2}
\arcsin\left(\frac{r_{\rm abs}}{r_{\rm orb}}\right) \ \text{yr},
\end{equation}
where $r_{\rm orb}$ is the orbital radius of the absorber, $M_6$ is the black hole mass in units of $10^6\,M_{\odot}$, and $r_{\rm abs}$ is the projected size of the absorber.

To test this scenario, we assume the absorber is located at a distance comparable to the UV/optical emitting region, $r_{\rm orb} \approx 40$--70 light-days, as suggested by the blackbody fits. Assuming a clump size an order of magnitude smaller than the orbital radius ($r_{\rm abs} \approx 0.1\,r_{\rm orb}$), and adopting the estimated black hole mass, we obtain a crossing time of $\sim$180--220 days. This timescale is broadly consistent with the duration of the possible non-detection interval, providing tentative support for a compact cloud orbiting within a larger-scale wind structure.

In the clumpy-absorber scenario, the assumed absorber distance is
comparable to the expected broad-line region (BLR) size estimated from
the empirical $R_{\rm BLR}$--$L_{5100}$ relation. Optically thick
clumps transiting across our line of sight could therefore in principle
also influence the ionizing continuum incident on the BLR. In practice,
the optical continuum and broad emission-line profiles remain remarkably
stable over decade-long timescales, indicating that the BLR is not
significantly affected during the X-ray low states. Furthermore, we estimate the Balmer decrement and obtain a value of
$\sim3$, comparable to the intrinsic broad-line ratio expected for
unobscured type~1 AGNs
($f_{\mathrm{H}\alpha}/f_{\mathrm{H}\beta} \approx 3$;
\citealt{Veilleux1987, Dong2008}), suggesting that the BLR is not
strongly affected by dust obscuration. Together, these results suggest
that the absorbing material either subtends only a limited covering
fraction as viewed from the BLR or is largely confined to our direct
line of sight toward the compact X-ray-emitting region.

Nevertheless, some aspects of the observed variability remain difficult to reconcile with a simple clumpy-absorber scenario. In particular, the rapid flux rise observed in September 2021 occurs on a much shorter
timescale ($\sim$12 days), which would require sharp column-density
gradients or highly structured absorbing material at the cloud
boundary. Such rapid variability may instead indicate that intrinsic
changes in the X-ray corona also contribute to the observed behavior.
Rapid variations in coronal geometry and energetics on timescales of
days have been proposed to explain extreme X-ray variability in AGNs
(e.g., \citealt{Ricci2020, Wu2020, Papoutsis2026, Xu2025}). Several
physical scenarios involving variable coronae have been suggested,
including ``failed jet'' configurations and extreme reprocessing models
(e.g., \citealt{Ghisellini2004, Lawrence2018}). Therefore, although a
clumpy absorber crossing our line of sight provides a natural
explanation for the long-term X-ray suppression, intrinsic coronal
variability cannot currently be ruled out. Future broadband X-ray
monitoring with improved spectral sensitivity will be important for
distinguishing between variable absorption and intrinsic coronal
variability in SDSS~J0005+2007.

\subsection{Comparison with X-ray--Weak Quasars}
X-ray--weak quasars constitute a population of AGNs whose X-ray
emission is significantly weaker than expected from their UV
luminosities. These objects typically lie below the canonical
$\alpha_{\rm ox}$--$L_{2500\,\mathring{A}}$ relation defined for normal
radio-quiet quasars \citep{Steffen2006}, often corresponding to X-ray
weakness factors of $f_{\rm weak} \gtrsim 10$. As shown by the spectral
energy distribution analysis presented in Section~\ref{sec:34},
SDSS~J0005+2007 undergoes a clear transition into an X-ray--weak state.

\begin{figure*}
 \begin{flushleft}
  \includegraphics[width=16cm]{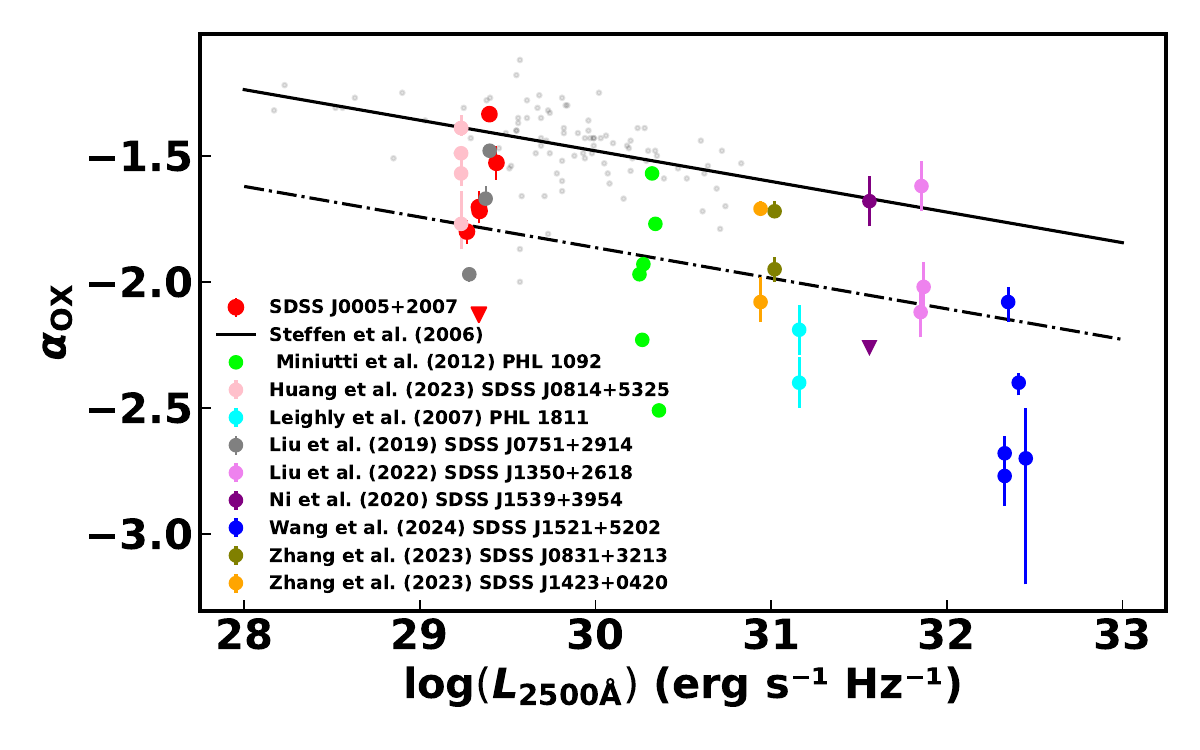} 
 \end{flushleft}
\caption{X-ray–to–optical slope ($\alpha_{\rm OX}$) versus the monochromatic luminosity at rest-frame 2500~\AA.
The red points show SDSS~J0005+2007 at different epochs.
Gray points represent typical AGNs from \citet{Steffen2006}, and the
solid black line shows their best-fit $\alpha_{\rm OX}$–$L_{2500\text{\AA}}$
relation. Large colored symbols denote individual X-ray weak quasars with multi-epoch observations, illustrating their variability trajectories. These include PHL~1092 \citep{Miniutti2012}, the prototype PHL~1811 \citep{Leighly2007}, SDSS~J0751+2914 \citep{Liu2019}, SDSS~J0814+5325 \citep{Huang2023}, SDSS~J1350+2618 \citep{Liu2022}, SDSS~J1539+3954 \citep{Ni2020}, SDSS~J1521+5202 \citep{Wang2024}, and the variable WLQs (SDSS~J0831+3213 and SDSS~J1423+0420) reported by \citet{Zhang2023}. The dot–dashed line marks an X-ray weakness factor of
$f_{\rm weak}=10$ ($\Delta\alpha_{\rm OX}=-0.384$).}\label{fig:aoxtend}
\end{figure*}

Figure~\ref{fig:aoxtend} places SDSS~J0005+2007 in the context of
several well-studied X-ray--weak quasars, including PHL~1811 and its
analogs. During the high-flux states, SDSS~J0005+2007 lies close to the
locus of typical quasars, with $\alpha_{\rm ox}$ values consistent with
the expected relation, indicating a normal disk--corona configuration.
However, during the low-flux epochs the source moves substantially
below the relation, entering the region populated by known X-ray--weak
quasars such as PHL~1811 \citep{Leighly2007} and PHL~1092
\citep{Miniutti2012}. The inferred $\Delta\alpha_{\rm ox}$ values and
X-ray weakness factors are comparable to those observed in these
systems, demonstrating that SDSS~J0005+2007 reaches a level of X-ray
weakness typical of this extreme population. This behavior suggests
that at least a fraction of X-ray--weak quasars may represent
transient episodes of X-ray suppression rather than intrinsically
distinct objects.

Similar transitions have been observed in several quasars. As prominent examples,
SDSS~J0751+2914 \citep{Liu2019}, SDSS~J0814+5325 \citep{Huang2023},
SDSS~J1350+2618 \citep{Liu2022}, and SDSS~J1539+3954 \citep{Ni2020}
all exhibit dramatic X-ray variability with little corresponding UV
change. In particular, SDSS~J1539+3954 shows an X-ray flux increase by
a factor of $\gtrsim 20$ while its UV continuum remained constant,
interpreted as the transverse motion of a geometrically thick inner
disk. More recently, \citet{Wang2024} reported a factor of $\sim 32$
variation in the ultraluminous quasar SDSS~J1521+5202 over 17~yr,
supporting models in which variable absorption regulates the observed
X-ray emission. In addition, \citet{Zhang2023} reported two X-ray--weak
quasars that returned to normal X-ray brightness, further indicating
that X-ray--weak phases can be transient.

SDSS~J0005+2007 also shares several properties with other X-ray--weak
AGNs. The relatively weak \ion{Mg}{2} emission and moderately high
Eddington ratio place it in a parameter space overlapping with
WLQs, many of which are X-ray weak. Moreover, the
extreme variability is confined primarily to the X-ray band, while the
UV and optical emission remain largely stable, a behavior commonly
observed in X-ray--weak quasars
\citep{Miniutti2012,Ni2020,Zhang2023,Huang2023}. 

The source also shows similarities to the NLS1 galaxy
RX~J0134.2$-$4258, which exhibits strong X-ray variability together
with weak high-ionization emission lines \citep{jin2023extreme}. In addition,
SDSS~J0005+2007 resembles the prototypical X-ray--weak NLS1
PHL~1092 \citep{Miniutti2012}, which shows dramatic X-ray dimming
without significant UV variability and stable optical spectra.
These similarities suggest that a subset of NLS1s may experience
episodes of extreme X-ray weakness, possibly related to complex and variable inner disk--corona structures.

\subsection{Could be a Tidal Disruption Event?}
The majority of X-ray-selected tidal disruption events (TDEs) demonstrate X-ray decay over a timescale spanning months to a year, a phenomenon attributed to the decline in accretion rates \citep{saxton_x-ray_2020}. While extremely rare, some TDEs exhibit X-ray emissions lasting up to a decade, with the notable exception of XJ1500+015 \citep{lin_likely_2017}. In contrast to typical TDE behavior, SDSS J0005+2007 displays persistent X-ray emissions for over 20 years, followed by a sharp decline over several years.

The emission line profiles in TDEs are known to evolve over time, with broad TDE lines typically narrowing as time progresses \citep{holoien_six_2016}. This change is attributed to the decreasing optical depth for electron scattering in the line emission region \citep{roth_what_2018}. However, the emission lines of SDSS J0005+2007 remain largely unchanged throughout the entire event.


Therefore, given the aforementioned considerations, the TDE scenario for SDSS J0005+2007 appears less likely.

\section{Summary}\label{sec:5}

We present a comprehensive multi-epoch study of
SDSS~J0005+2007, a quasar that transitions from an
X-ray--normal state to an X-ray--weak state. Over a
timescale of $\sim5$~yr, the 0.2--10~keV X-ray flux
declines by more than an order of magnitude, reaching
weakness factors of $f_{\rm weak} \gtrsim 10$--100 in
the faintest epochs. During the high states, the X-ray
spectrum is persistently soft, whereas the stacked
low-state spectrum shows substantial flux suppression
together with indications of spectral hardening.

In contrast, the UV continuum varies only mildly
($\sim0.3$~mag), and the optical continuum, broad
emission lines, and mid-infrared emission remain stable
over decade-long timescales. The source moves from the
canonical $\alpha_{\rm ox}$--$L_{2500\,\mathring{A}}$
relation into the X-ray--weak regime during the low
states, while retaining the optical spectroscopic
characteristics of a normal type~1 quasar.

The pronounced wavelength dependence of the variability
provides strong physical constraints. The stability of
the optical continuum and broad-line region indicates
that the accretion disk structure and photoionizing
luminosity remain largely intact, whereas the dramatic
suppression of the X-ray emission must originate in the
compact coronal region or along our line of sight to it.
The spectral hardening observed in the low state, together
with the viability of ionized partial-covering models,
is consistent with an absorption-driven scenario in
which variable, largely dust-free gas located interior
to or comparable to the broad-line region modulates the
observed X-ray emission. Such behavior is similar to
that observed in several X-ray--weak quasars and is
consistent with clumpy inner disk winds or shielding-gas
configurations.

While intrinsic coronal variations cannot be excluded,
the multi-wavelength constraints disfavor a global
accretion-state transition. Given the specific observed 
characteristics, the possibility of a TDE scenario for SDSS J0005+2007 is low.
SDSS~J0005+2007 therefore
provides further evidence that extreme X-ray weakness
can arise as a transient phase in otherwise normal
quasars. Continued monitoring will be essential to
constrain the geometry and location of the absorbing
material and to determine the recurrence timescale of
the X-ray--weak phase.

\begin{acknowledgments}
We acknowledge the supports from National Natural Science Foundation of China (NSFC; grant Nos.12573110, 12133001), the Shenzhen Science and Technology Program (JCYJ20230807113910021) and the Natural Science Foundation of Top Talent of SZTU(GDRC202208).

Spectroscopic observations were obtained with the Palomar 200-inch Hale Telescope at Palomar Observatory using the Next Generation Palomar Spectrograph (\href{https://www.astro.caltech.edu/ngps}{NGPS}). We thank the Palomar Observatory staff for their support during the observations.
This work makes use of observational data obtained with Einstein Probe, a
space mission supported by Strategic Priority Program on Space Science of Chinese Academy of Sciences, in collaboration with ESA, MPE and CNES (Grant
No. XDA15310000), the Strategic Priority Research Program of the Chinese
Academy of Sciences (Grant No. XDB0550200), and the National Key R\&D
Program of China (2022YFF0711500).

\end{acknowledgments}

\end{document}